\documentclass[runningheads]{svmult}

\usepackage{makeidx}   
\usepackage{graphicx}  
\usepackage{subeqnar}  
\usepackage{multicol}  
\usepackage{physprbb}  
\makeindex             
\begin{document} 
\title*{Heavy Mesons and Impact Ionization of Heavy Quarkonia
\footnote{Published in: Lect. Notes Phys. 647:366-375,2004}} 
\toctitle{Heavy Mesons and Impact Ionization of Heavy Quarkonia} 
%
%
\titlerunning{Heavy Mesons and Impact Ionization} 
%
\author{David Blaschke\inst{1,2}
\and Yuri Kalinovsky\inst{3}
\and Valery Yudichev\inst{2}
} 
\authorrunning{David Blaschke et al.} 
%
%
\institute{Fachbereich Physik, Universit\"at Rostock, D-18051 Rostock, 
Germany 
\footnote{Presently at: Fakult\"at f\"ur Physik, Universit\"at Bielefeld,
D-33615 Bielefeld, Germany}
\and Bogoliubov Laboratory for Theoretical Physics, 
     Joint Institute for Nuclear Research, 
     141980 Dubna, Russia 
\and Laboratory of Information Technologies, 
    Joint Institute for Nuclear  Research, 
    141980 Dubna, Russia
    } 
 
\maketitle              
 
\begin{abstract} 
At the chiral restoration/deconfinement transition, most hadrons 
undergo a Mott transition from being bound states in the confined phase to 
resonances in the deconfined phase. We investigate the consequences of 
this qualitative change in the hadron spectrum on 
final state interactions of charmonium in hot and dense matter, and show that 
the Mott effect for D-mesons leads to a critical enhancement of the J/$\psi$ 
dissociation rate. Anomalous J/$\psi$ suppression in the NA50 experiment is 
discussed as well as the role of the Mott effect for the heavy flavor kinetics 
in future experiments at the LHC. The status of our calculations of 
heavy quarkonium dissociation cross sections due to quark and gluon impact 
is reviewed, and estimates for in-medium effects due to the lowering of the
ionisation threshold are given. 
\end{abstract}

\section{Introduction}
Heavy quarks as constituents of heavy quarkonia and heavy mesons play a
decisive role in the diagnostics of hot and dense matter created in 
ultrarelativistic heavy-ion collisions. The most promiment example is
the celebrated J/$\psi$ suppression effect as a possible signal of
the formation of a \index{quark-gluon plasma} which was suggested by 
Matsui and Satz \cite{Matsui:1986dk} in 1986 and has triggered experimental 
programs 
at CERN, Brookhaven and the future GSI facility as well as a broad spectrum 
of theoretical work ever since.
In the present lecture, we will elucidate some aspects of the physics of
heavy quarks in a hot and dense medium which demonstrate why J/$\psi$
suppression is so difficult to interprete. We suggest that $\Upsilon$ may
be a rather clean probe of the plasma state to be produced at CERN-LHC.

The production of heavy quark pairs is a hard process well separated from
the soft physics governing the evolution of a quark gluon plasma in 
heavy-ion collisions. However, the formation of heavy quarkonia and heavy
mesons as well as their final state interactions may be well modified by 
a surrounding hot medium. The basic principle of their use as indicators
is similar to the situation in Astrophysics where information about the 
temperature, density, composition and other properties of stellar atmospheres
is obtained from an analysis of the modification of emission and absorption
spectra relative to terrestrial conditions \cite{Kajantie:bs}.
 
The first and widely discussed probe in this context is the J/$\psi$.
Contrary to naive expectations, Matsui and Satz suggested that with increasing
cms energy in a heavy-ion collision a suppression of the   J/$\psi$
production relative to the (Drell-Yan) continuum occurs due to the 
dissociation of the charmonium bound state in a dense medium.
This is in analogy to the \index{Mott transition} in semiconductor and 
plasma physics where under high pressure electrons become delocalized and a 
conduction band emerges, signaling the plasma state \cite{rr}. 

Correlators of quarkonia states are measured at finite temperature on the 
lattice \cite{asakawa99,asakawa01,umeda01,umeda02,datta02}. 
Using the maximum entropy method \cite{asakawa99}, spectral functions can 
be obtained which give
not only information about the mass but also about the spectral width of the 
states.
No substantial modification of J/$\psi$ spectral function is found for 
temperatures up to $T=1.5 T_c$.
The $\chi_c$ and $\psi'$ states may be dissolved at $T_c$.
This new lattice analysis provides results alternative to those obtained 
earlier by solving the bound state Schr\"odinger equation for a 
(temperature-dependent) screened potential \cite{karsch88}.
While the Mott-dissociation of   $\chi_c$ and $\psi'$ states is confirmed,
the observation of a nearly unchanged J/$\psi$ resonance well above the 
suspected Mott temperature $T^{\rm Mott}_{J/\psi}=1.1-1.3~T_c$ came as a 
surprise for simple potential models.
The existence of \index{resonances} above the Mott transition is, however, a 
well-known feature displayed, e.g., in analyses of exciton lines in 
semiconductor plasmas \cite{rr}.
The picture which emerges from these lattice studies seems to be consistent 
with a modification of the effective interaction at finite temperatures so
that excited states of the charmonium spectrum  ($\chi_c$ and $\psi'$) can
be dissolved at $T_c$ whereas the deeply bound ground states ($\eta_c$ and 
J/$\psi$) are observed as rather narrow resonances well above that temperature.
In order to interpret these findings a study of the spectral functions for
screened Coulomb or Cornell-type potentials should be performed.

\begin{figure}
\centerline{\includegraphics[width=2.5in,angle=0]{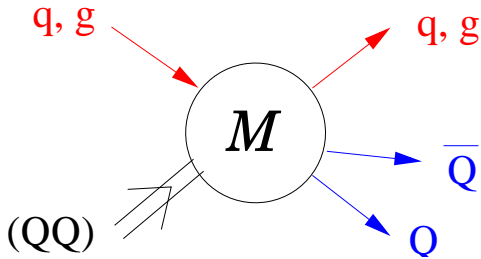}}
\caption{Transition amplitude for a quarkonium breakup process in a 
fully developed quark-gluon plasma.}
\label{breakup2}
\end{figure}

\section{Quantum kinetics for quarkonium in a plasma}

\begin{figure}
\includegraphics[width=4in,angle=0]{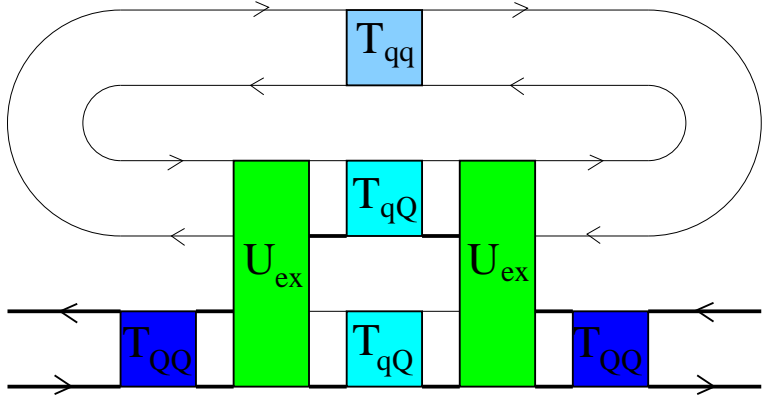}
\caption{Transition amplitude for heavy quarkonium dissociation
in strongly correlated quark matter. The T-matrices stand for mesonic 
bound and/or scattering states. }
\label{sigma-h}
\end{figure}

The {inverse lifetime} of a state with 4-momentum $p$ in a plasma is 
related to the imaginary part of its selfenergy by 
$\tau^{-1}(p) = \Gamma(p) = \Sigma^>(p) - \Sigma^{<}(p)$~,
where in the {Born approximation for} the quarkonium breakup
by quark and gluon impact holds, see also Fig. \ref{breakup2}
\begin{eqnarray}
\Sigma^{\stackrel{>}{<}}(p)&=& \int_{p'}\int_{p_1}\dots \int_{p_3} (2\pi)^4 
\delta_{p+p',p_1+p_2+p_3} |{\cal M}|^2
G^<(p') G^>(p_1) G^>(p_2) G^>(p_3)~,
\nonumber
\end{eqnarray}
where the nonequilibrium Green functions can be expressed via
distribution functions $f_i(p)$ and spectral functions $A_i(p)$
of the particle species $i$ as
$ 
G^>_i(p)= [1\pm f_i(p)],~G^<_i(p)= f_i(p) A_i(p)~.
$
This quantum kinetic formulation includes not only the loss but also the 
reverse process of $Q\bar Q$ fusion in a consistent way.
It can also be generalized to include strong correlations such as hadronic
bound states or resonances in the visinity of the hadronization transition,
see Fig. \ref{sigma-h}.
First steps in this direction have been explored \cite{Burau:2000pn}.

\section{Quarkonium dissociation cross section in a gluon gas}

In this section we give benchmark results for activation of 
Coulombic bound states by collisions in a medium. The medium will be 
represented as a gluon gas with thermal distribution function,
$n_g(\omega)=g_g [\exp(\omega(p)/T) - 1]^{-1}$, where
the degeneracy factor $g_g=2 (N_c^2-1)$.
\begin{table}[htb]
\begin{tabular}{|c|cc|cc|}
\hline
&set(i)&&set(ii)&\\ 
\hline
$Q \bar{Q}$ system & $\varepsilon_0 $[GeV] & $m_Q [GeV]$ & 
$\varepsilon_0$ [GeV] & $m_Q$ [GeV]\\
\hline
bottomonium & 0.75 & 5.10 & 1.10 & 5.28 \\
charmonium & 0.78 & 1.94 & 0.62 & 1.86 \\
\hline
\end{tabular} 
\caption{Parameters of the heavy quarkonium systems: binding energy 
$\varepsilon_0$ and heavy quark mass $m_Q$ from \cite{arleo}. Left two 
columns refer to set(i), the right two columns to set(ii).}
\label{tbl.hqp}
\end{table}

\begin{figure}
\centerline{\includegraphics[width=4in,angle=-90]{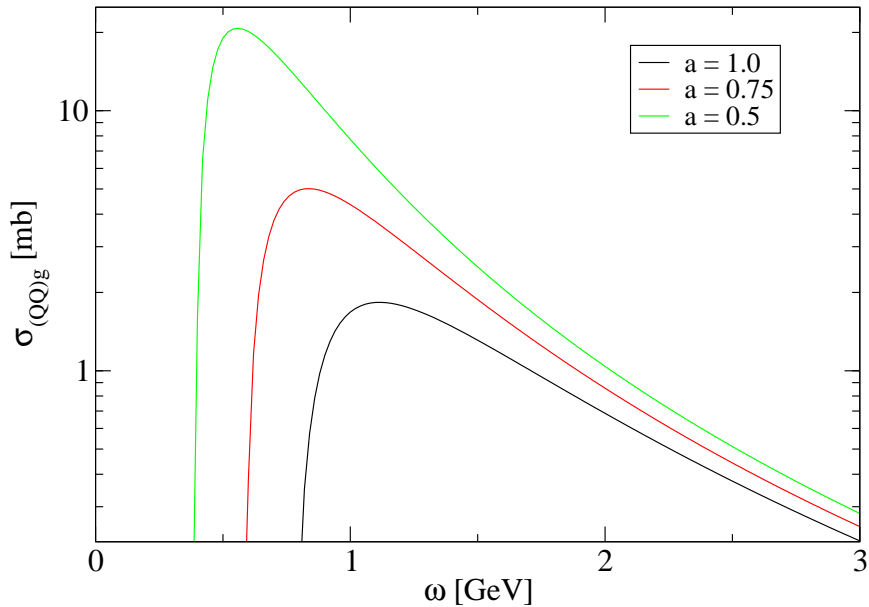}}
\caption{Energy dependence of the quarkonium breakup cross section by 
gluon impact for different (constant) threshold depletion factors 
$a(T)=1, 0.75, 0.5$.}
\label{sigma}
\end{figure}

The quarkonium breakup cross section by gluon impact
can be estimated with the Bhanot/Peskin formula \cite{bp}
\begin{equation}
\sigma_{(Q\bar Q) g} (\omega)= 
\frac{2^{11}}{3^4}\alpha_s \pi a_0^2
\frac{(\omega/\epsilon_0 - a(T))^{3/2}}{(\omega/\epsilon_0)^5} 
\Theta(\omega - \epsilon_0a(T))
\end{equation}
with the binding energy $\epsilon_0$ of the $1S$ 
quarkonium state with a Coulombic rms radius 
$\sqrt{\langle r^2\rangle_{1S}}=\sqrt{3} a_0=
2\sqrt{3}/(\alpha_s m_Q)$, and the heavy-quark mass $m_Q$, the energy of the 
impacting gluon is $\omega$. 
With the parameter $0<a(T)<1$ we can vary the effective dissociation threshold 
from $\omega_{th}^{ideal}=\epsilon_0; a(T<T_c)=1$ to  
$\omega_{th}^{Mott}=0; a(T>T^{Mott})=0$. 
See Fig. \ref{sigma} for numerical examples.  

\begin{figure}
\centerline{\includegraphics[width=4in,angle=-90]{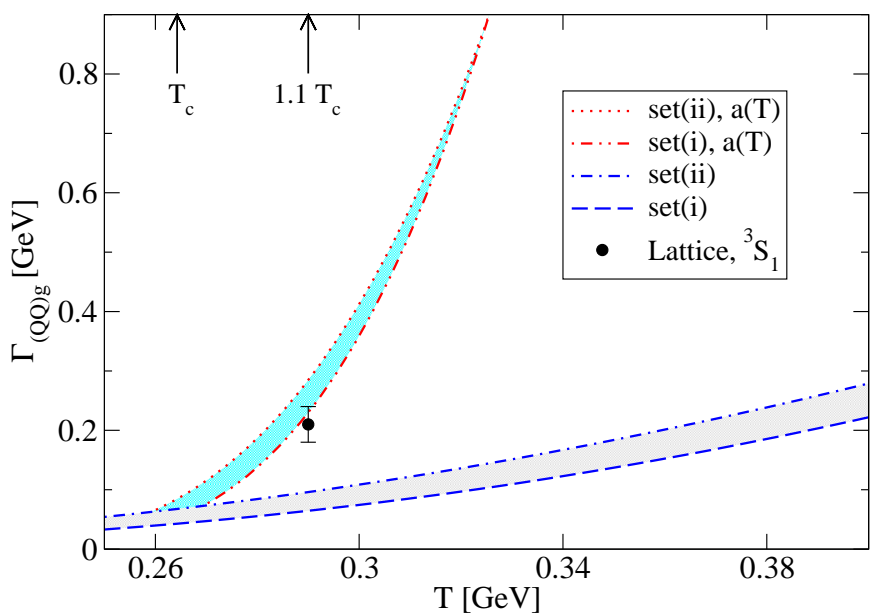}
}
\caption{Rate coefficient $\Gamma_{(Q\bar Q)g}=\tau^{-1}$
for heavy quarkonium dissociation by gluon impact as a function of the
gluon plasma temperature (no medium, $a(T)=1$),
parameters see Tab. \ref{tbl.hqp}.
For comparison, the Breit-Wigner fit of the spectral width for $^1S_0$ 
charmonium from lattice simulations is shown.
In-medium effects are estimated by a lowering of the breakup threshold,
see text.}
\label{gamma-tau-g}
\end{figure}

The dissociation rate for a heavy quarkonium $1S$
state at rest in a heat bath of massless gluons ($\omega=p$)
at temparature $T$ is
\begin{equation}
\frac{1}{\tau_{(Q\bar Q) g}(T)}=\Gamma_{(Q\bar Q) g}(T)=
\langle \sigma_{(Q\bar Q) g}^{ideal}(\omega)n_g(\omega)\rangle_T
= \frac{1}{2\pi^2}\int_0^\infty \omega^2 d \omega 
\sigma_{(Q\bar Q) g}^{ideal}(\omega)n_g(\omega)
\end{equation}
and is shown in Fig. \ref{gamma-tau-g}. 
  
It is interesting to note that there is a discrepancy between the 
Lattice calculation of the charmonium decay with in a gluonic medium and the
decay width of charmonium by gluon impact due to the gluonic E1 transition. 
It is conceivable that the Lattice simulation overestimates the 
thermal width of the charmonium states, but we would rather like to suggest
that due to a lowering of the ionisation threshold in the 
gluonic medium the breakup rate will be enhanced. 
For an analytical estimate we adopt a temperature dependence of the strong 
coupling constant by using the gluon momentum scale \cite{kapusta}
$\langle p^2\rangle = (3.2 ~T)^2$
in the 1-loop $\beta$ function such that 
\begin{equation}
\label{alpha_T}
\frac{\varepsilon_0~a(T)}{m_Q}=
\left(\frac{3\pi}{22~\ln(3.2~T/\Lambda)}\right)^2~.
\end{equation}
\vspace{-1cm}

\begin{figure}
\centerline{\includegraphics[width=4in,angle=-90]{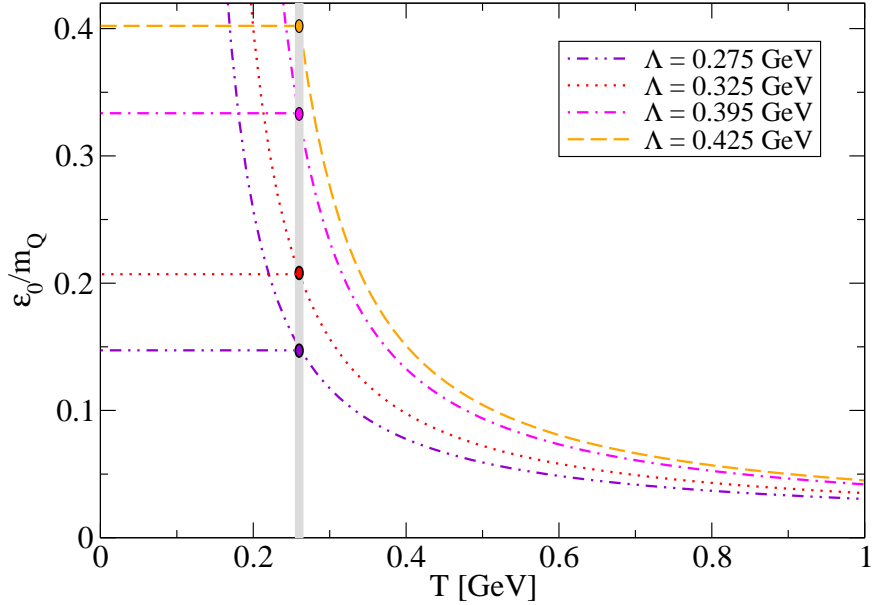}
}
\caption{
Temperature dependence of the ionisation threshold due to the 1-loop
running of the strong coupling constant (\ref{alpha_T}).}
\label{alpha-T}
\end{figure}

\section{Quarkonia abundances and observable signatures}

\begin{figure}
\includegraphics[width=4in,angle=-90]{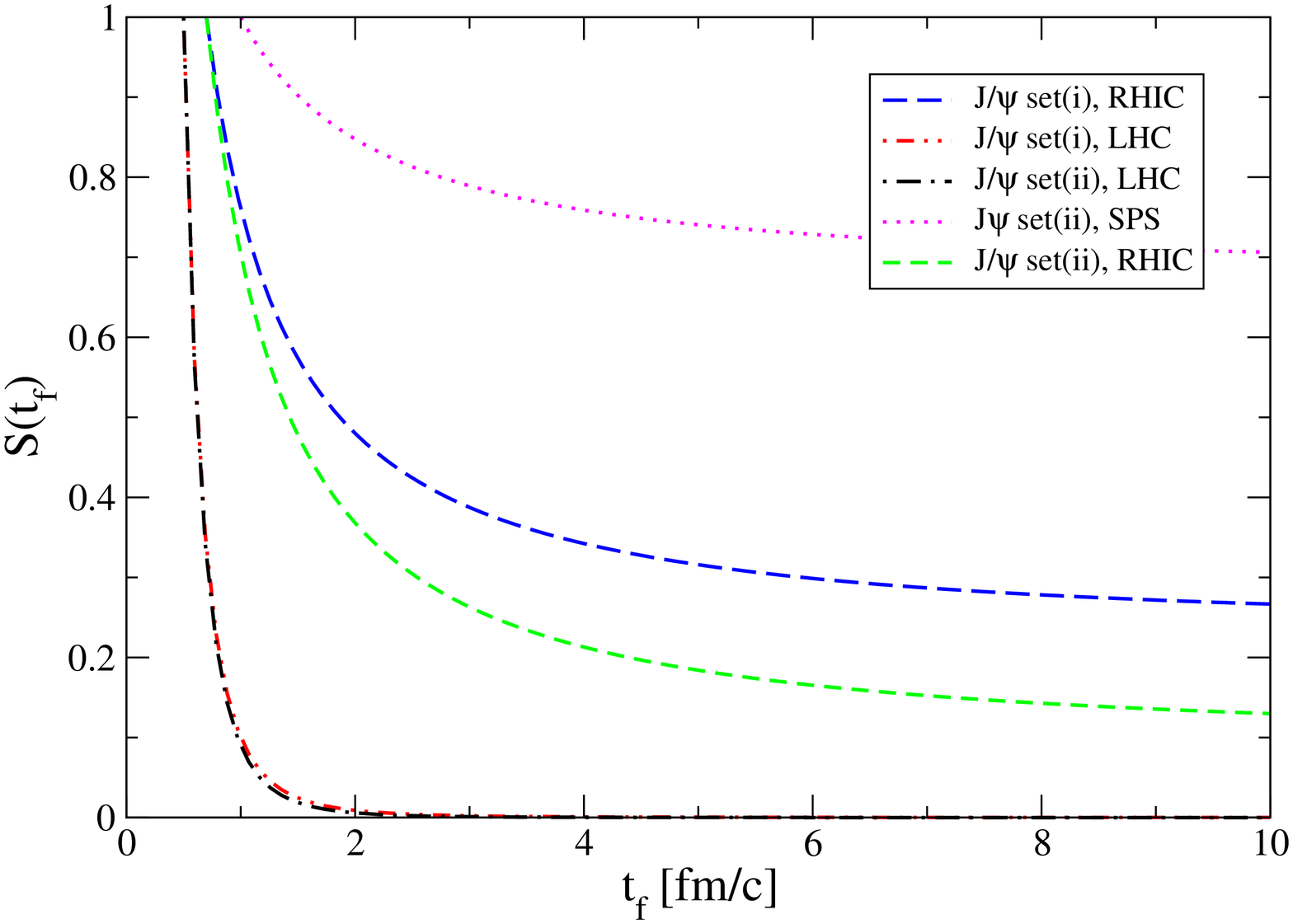}
\includegraphics[width=4in,angle=-90]{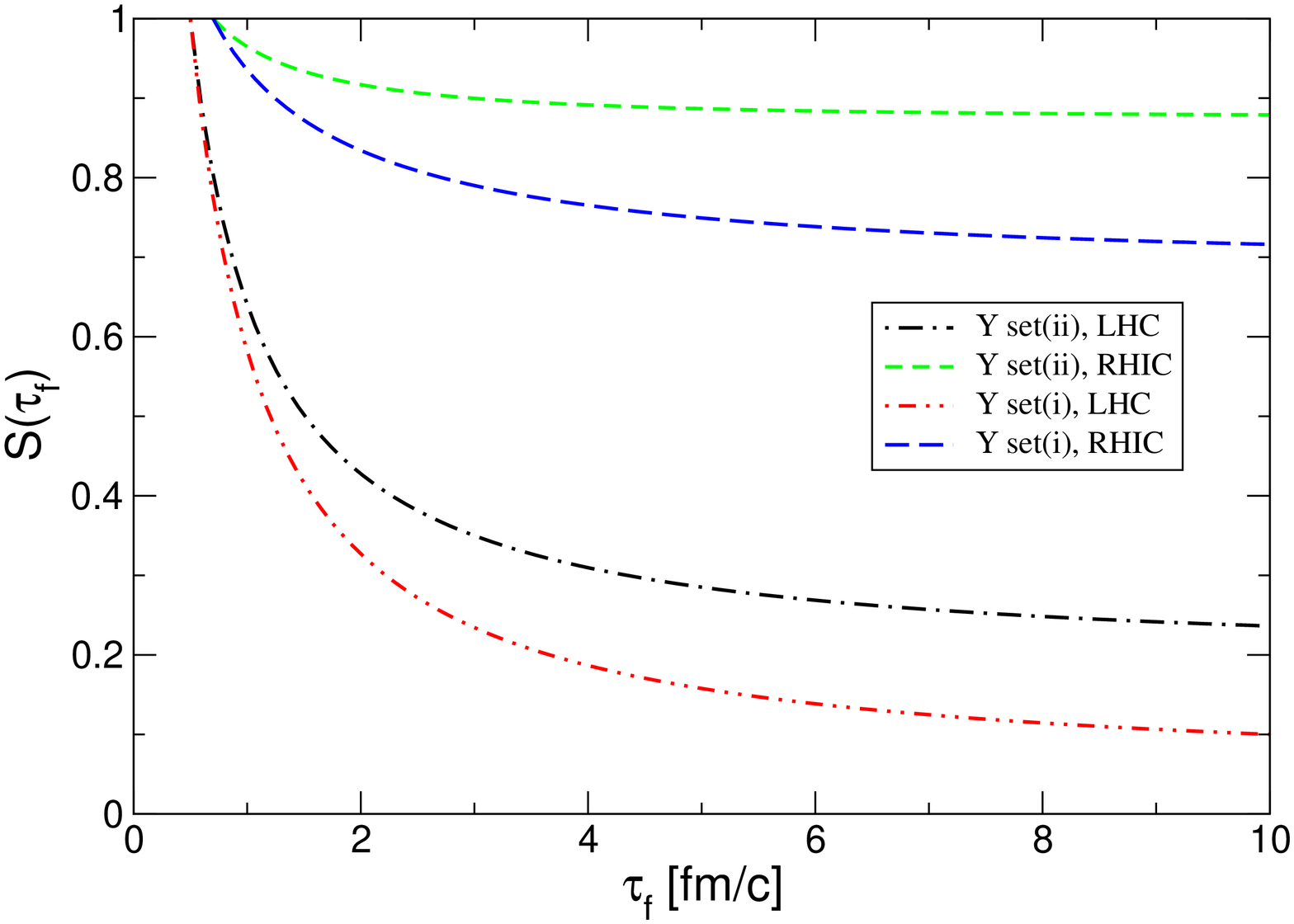}
\caption{Survival probability for heavy quarkonia states in a 
longitudinally expanding gluon plasma as a function of the plasma lifetime;
$^1S_0$ charmonium (J/$\psi$, upper panel) and $^1S_0$ 
bottomonium ($\Upsilon$, lower panel), parameters see Tabs. 
\ref{tbl.hqp}, \ref{tbl.HIC}.}
\label{suppr}
\end{figure}

In order to study observable signatures we will adopt here the Bjorken 
scenario \cite{Bjorken:1982qr} for the plasma evolution, i.e. longitudinal expansion with conserved entropy: $T^3~t=T_0^3~t_0={\rm const}$.
Parameters for the initial state are given in Tab. \ref{tbl.HIC}, where
RHIC and LHC values are taken from \cite{xu}, the SPS values are 
''canonical''. 
\begin{table}[htb]
\begin{tabular}{cccc} 
\hline
 & LHC (3) & RHIC (3) & SPS  \\
\hline
$T_0$ [GeV] & 0.72 & 0.4 & 0.25 \\
$\tau_0$ [fm/c] & 0.5 & 0.7 & 1.0 \\
\hline
\end{tabular} 
\caption{Heavy-ion collision parameters from \cite{xu}.}
\label{tbl.HIC}
\end{table}
As the quantity for comparison with experimental data of quarkonium production 
we consider the survival probability. We neglect here the hadronic comover and 
the nucleonic contributions, and also the effects due to the hadronisation 
phase transition
\begin{equation}
S(t_f)=\exp\left(-\int_{t_0}^{t_f}~dt~\tau^{-1}(T) \right)~.
\end{equation}
At the freeze-out time $t_f$ the collisions stop to change the number of 
J/$\psi$ ($\Upsilon$). Using the Bjorken scaling it can be related into
a freeze-out temperature. In Fig. \ref{suppr} we give the survival probability
for J/$\psi$ (left panel) and for $\Upsilon$ (right panel) for the parameters
of the Tables \ref{tbl.hqp} and \ref{tbl.HIC} due to gluon impact.
For the LHC conditions, the simple estimates presented here don't give robust
predictions for J/$\psi$. The $\Upsilon$, however, shall be a good probe for
the lifetime of the plasma state as well as for its temperature. 
More details can be found in Ref.  \cite{Bedjidian:2003gd}.

\section{Quarkonia dissociation by quark impact, $T>T_c$}
\begin{figure}
\includegraphics[width=4in,angle=-90]{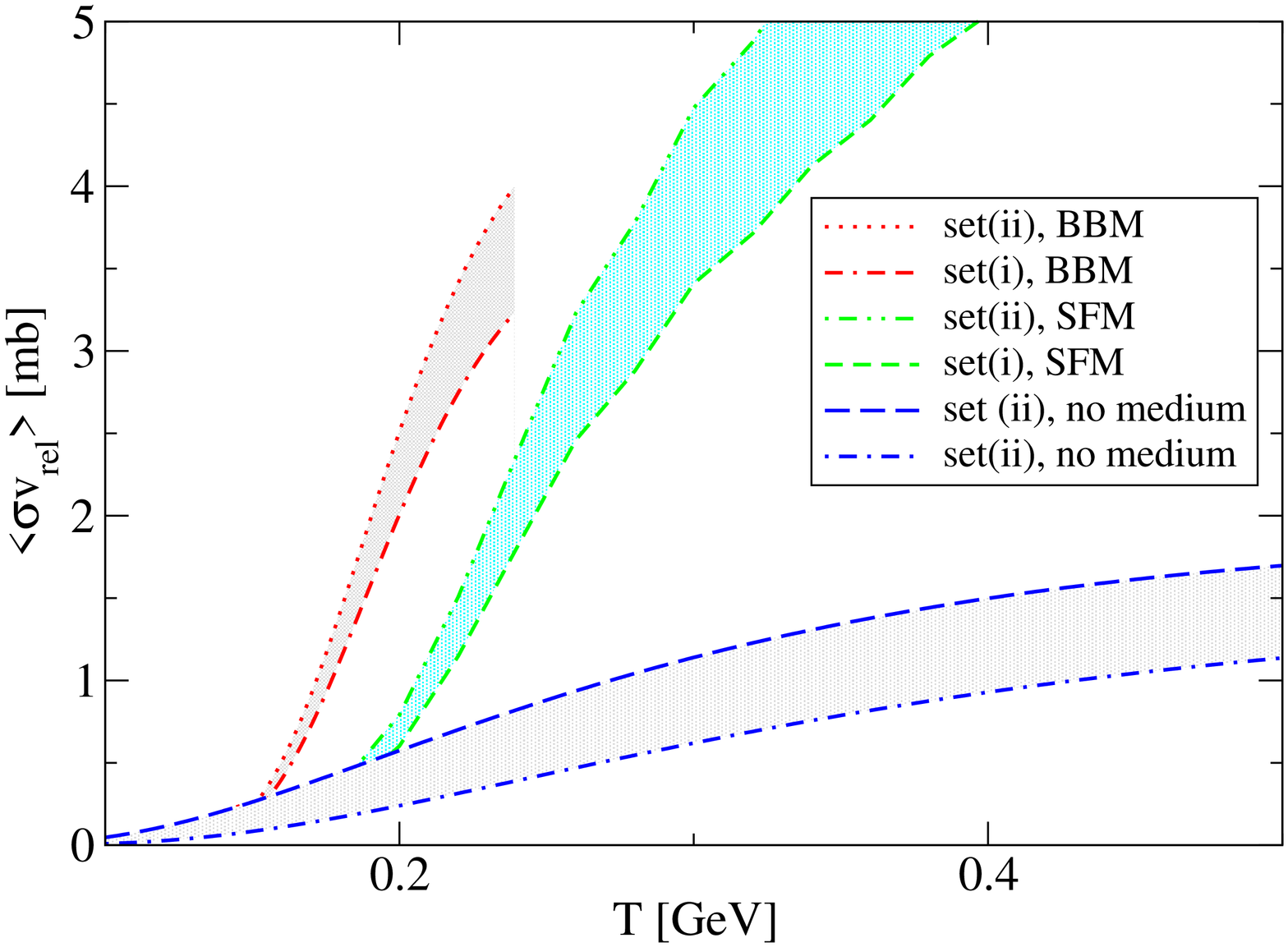}
\includegraphics[width=4in,angle=-90]{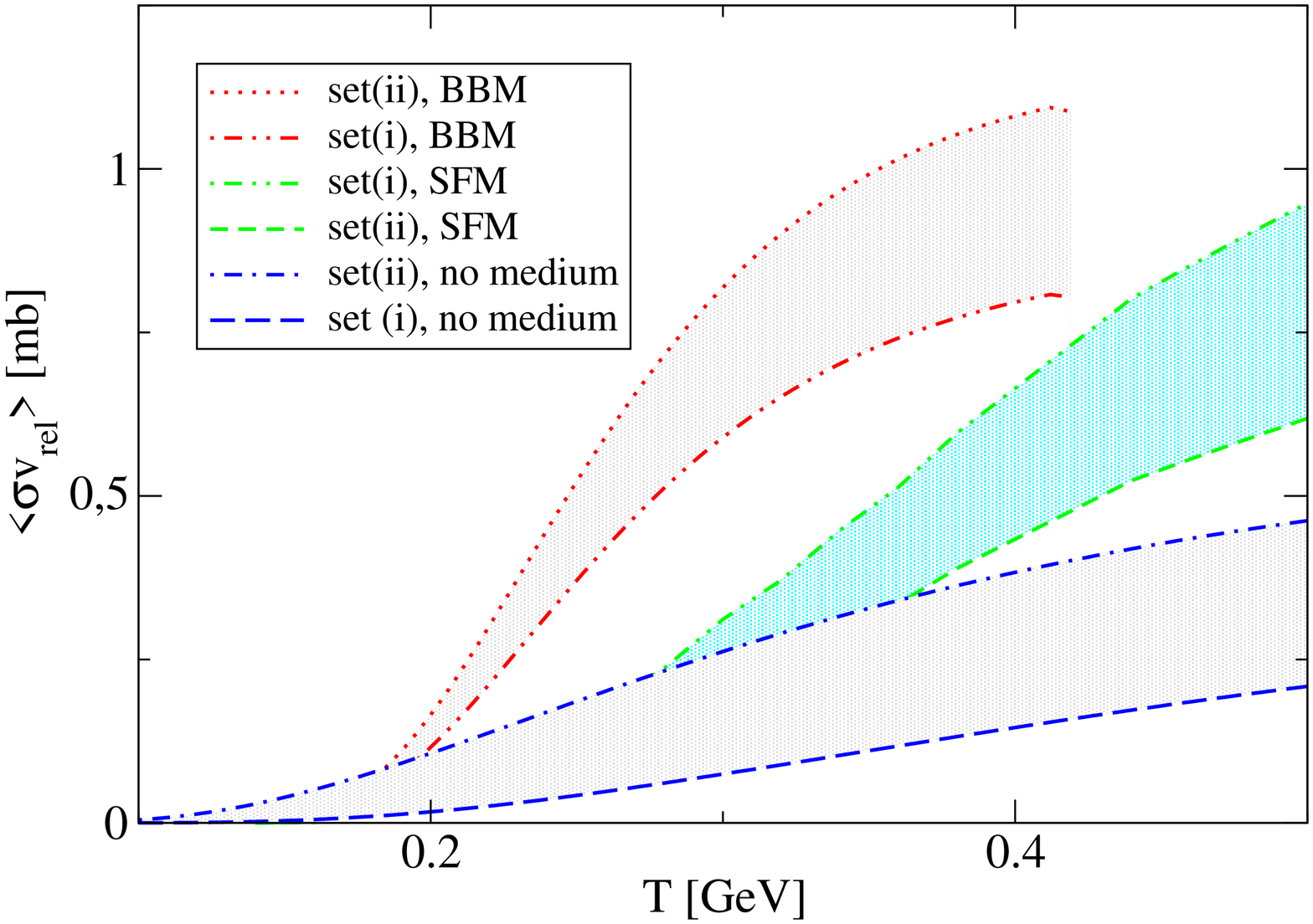}
\caption{Thermally averaged cross section for heavy quarkonium dissociation
by quark impact as a function of the temperature: J/$\psi$ (upper panel) 
and $\Upsilon$ (lower panel), parameters see Tab. \ref{tbl.hqp}.
BBM and SFM approach give a similar cross section enhancement due to 
the lowering of the breakup threshold.}
\label{sigma-psi-ups}
\end{figure}

The quarkonium breakup cross section by quark impact is estimated
using the Bethe formula \cite{bethe} for impact ionization of 
the $1S$ bound state of the Coulomb potential  
(the quark now plays the role of the impacting electron)
\begin{equation}
\sigma_{(Q\bar Q) q} (\omega)= 2.5 \pi a_0^2
\frac{\varepsilon_0}{\omega}
\ln\left(\frac{\omega+\Delta(T)}{\varepsilon_0}\right)
 \Theta\left(\frac{\omega + \Delta(T)}{\varepsilon_0}-1\right)~.
\end{equation}
Here we employ a generalization of Bethe's formula which takes into account
the shift of the continuum edge of scattering states by $\Delta(T)$, which
results in a temperature-dependent lowering of the ionisation threshold
\cite{schlanges}, $\varepsilon(T)=\varepsilon_0-\Delta(T)$. This is named
the Bethe-Born model (BBM).
For quark matter systems, the energy shift $\Delta(T)$ has been obtained
within the string-flip model (SFM), see \cite{Blaschke:1984yj,Ropke:qs}.
Within this model also an effective cross section for heavy quarkonium
breakup has been estimated \cite{rbs,rbs88}
\begin{equation}
\sigma(T)=\pi r^2_{Q\bar Q}(T)\exp(-\Delta (T)/T)
\end{equation}
It is remarkable that the enhancement of the breakup cross section due
to the lowering of the threshold in the BBM is comparable in magnitude 
to that of the SFM approach, see Fig. \ref{sigma-psi-ups}.

\section{Conclusions}
In the QGP (and in the mixed phase), due to the presence of quasifree quarks
and gluons, new channels for charmonium formation and dissociation exist which
could drive chemical equilibration during the
existence of the fireball formed in the heavy-ion collision.
In this contribution we have given some benchmark estimates which have to
be further elaborated.
The role of partonic in-medium effects in the heavy quarkonium kinetics 
in a QGP, which has previously been discussed in the string-flip model of 
quark matter in the
form of a modified mass action law \cite{rbs} and dissociation rate
\cite{rbs88}, should be reconsidered by also including gluonic degrees of
freedom which shall become dominant at RHIc and LHC conditions.
We suggest that rate coefficients for the ionization and recombination
of charm mesons could be described using an approach similar to methods
used previously to study Coulomb plasmas \cite{rr,bornath,schlanges}.
It is suggested that the measurement of the $\Upsilon$ suppression at LHC
can provide rather robust informations about the initial temperature
(or lifetime) of the quark-gluon plasma.

\section*{Acknowledgement}
V.Yu. acknowledges support by DFG under grant No. 436 RUS 17/37/03
and the Russian Fund for Fundamental Research under grant No. 02-02-16194
as well as by the Heisenberg-Landau program.
D.B. is grateful to many colleagues for numerous enlightening discussions,
in particular to G. Burau, J. H\"ufner, P. Petreczky, A. Polleri, H. Satz 
and  R. Vogt.

\end{document}